# Probing Perfection: The Relentless Art of Meddling for Pulmonary Airway Segmentation from HRCT via a Human-AI Collaboration Based Active Learning Method

Shiyi Wang[b,*], Yang Nan[a,*], Sheng Zhang[b], Federico Felder[c], Xiaodan Xing[a], Yingying Fang[b,c], Javier Del Ser[e,f], Simon L F Walsh [b,c,**], Guang Yang[a,b,c,d,**]

[a.] Bioengineering Department and Imperial-X, Imperial College London, London W12 7SL, UK;
[b.] National Heart and Lung Institute, Imperial College London, London SW7 2AZ, UK;
[c.] Cardiovascular Research Centre, Royal Brompton Hospital, London SW3 6NP, UK;
[d.] School of Biomedical Engineering & Imaging Sciences, King's College London, London WC2R 2LS, UK;
[e.] TECNALIA, Basque Research and Technology Alliance (BRTA), Derio 48160, Spain
[f.] Department of Communications Engineering, University of the Basque Country UPV/EHU, Bilbao 48013, Spain
[*] Co-first authors, [**] Co-last authors.

**Abstract.**

In the realm of pulmonary tracheal segmentation, the scarcity of annotated data stands as a prevalent pain point in most medical segmentation endeavors. Concurrently, most Deep Learning (DL) methodologies employed in this domain invariably grapple with other dual challenges: the inherent opacity of 'black box' models and the ongoing pursuit of performance enhancement. In response to these intertwined challenges, the core concept of our Human-Computer Interaction (HCI) based learning models (RS_UNet, LC_UNet, UUNet and WD_UNet) hinge on the versatile combination of diverse query strategies and an array of deep learning models. We train four HCI models based on the initial training dataset and sequentially repeat the following steps 1-4: (1) Query Strategy: Our proposed HCI models selects those samples which contribute the most additional representative information when labeled in each iteration of the query strategy (showing the names and sequence numbers of the samples to be annotated). Additionally, in this phase, the model selects the unlabeled samples with the greatest predictive disparity by calculating the Wasserstein Distance, Least Confidence, Entropy Sampling, and Random Sampling. (2) Central line correction: The selected samples in previous stage are then used for domain expert correction of the system-generated tracheal central lines in each training round. (3) Update training dataset: When domain experts are involved in each epoch of the DL model's training iterations, they update the training dataset with greater precision after each epoch, thereby enhancing the trustworthiness of the 'black box' DL model and improving the performance of models. (4) Model training: Proposed HCI model is trained using the updated training dataset and an enhanced version of existing UNet. Experimental results validate the effectiveness of this Human-Computer Interaction-based approaches, demonstrating that our proposed WD-UNet, LC-UNet, UUNet, RS-UNet achieve comparable or even superior performance than the state-of-the-art DL models, such as WD-UNet with only 15% - 35% of the training data, leading to substantial reductions (65%-85% reduction of annotation effort) in physician annotation time.

## 1 Introduction

### 1.1 Challenges

In the field of clinical health care, high-resolution computed tomography (HRCT) of the chest is routinely performed in most patients with suspected airway-related diseases. Such conditions include congenital tracheal stenosis, which causes airway obstruction[1] and chronic obstructive pulmonary disease (COPD) and asthma, all of which can lead to abnormal airway dilation[2]. The characteristics of pulmonary fibrosis on the tracheal structure can manifest as a thickening of the trachea and abnormal dilation at the ends of the airways.

Yet, traditional visual quantification on HRCT involving domain experts lead to inaccurate diagnoses. This is because the outcomes of visual quantification are based on the subjective estimations of radiologists, which are approximate values with variability. Thus, visual quantification of airway disease on HRCT is liable to high levels of inter-reader variability and has limited sensitivity to changes in disease severity over short follow-up periods[3]. In contrast, quantitative CT (QCT), which employs computer-based techniques for analyzing HRCT image segmentation predictions, offers an objective and reproducible alternative for evaluation, aiding radiologists in making more accurate diagnoses. In the lung CT images, the trachea exhibits specific visual characteristics and annotation standards, with generally minimal variability. Thus, accurate annotation of the trachea can effectively mitigate inter-reader variability caused by visual quantification. Consequently, with QCT acting as a downstream task of segmentation, the accuracy of airway segmentation models predictions directly influences the precision of quantification, impacting radiologists' diagnostic and prognostic assessments[4].

A precise tracheal segmentation model can extract the trachea from CT scans of patients with pulmonary fibrosis, thus facilitating diagnosis and prognosis. However, tracheal segmentation faces three main challenges: the need for extensive annotation support, the need for models with better predictive performance, and the "Black Box" nature of Deep Learning (DL) segmentation models. For instance, the method for extracting the tracheal centerline, as proposed by Lee[5], employs an octree data structure to examine a 3x3x3 pixel neighborhood. However, this approach frequently encounters challenges such as inaccurate extraction of the centerline, disorganized segmentation, and the emergence of looping patterns. Due to the objectives of Deep Human-Computer Interaction (HCI) approaches, which are to complement, empower, and in no way replace human capabilities, and considering that the task of annotating tracheal samples is inherently challenging and cannot be entirely substituted by automated methods, medical experts' domain-specific knowledge plays a crucial role in rectifying inaccuracies in machine-generated results.

### 1.2 Contributions

In our study, we used 3D UNet[5] and 3D CEUNet[6] as supervised baselines, and proposed four semi-supervised models: LC-UNet, UUNet, RS-UNet, and WD-UNet, based on various strategies like Least Confidence[7], Entropy Sampling[8], Random Sampling and Wasserstein Distance[9], which can save the heavy manual annotation cost and alleviate the label scarcity issue.

To overcome the aforementioned three challenges, we propose four Human-Computer Interaction-based learning models which involve domain experts annotating the trachea in pulmonary CT scans. Since the trachea has specific visual characteristics and annotation standards in pulmonary CT images, generally, there are no significant discrepancies[10]. Therefore, to make the results more trustworthy, semi-supervised airway segmentation model of lung fibrosis was achieved by involving radiologists in every epoch of DL model training process while incorporating deep learning (DL) into the annotation framework. Additionally, our proposed Human-Computer Interaction-based models – Wasserstein Distance-based UNet (WD-UNet), Least Confidence-based UNet (LC-UNet), Uncertainty-based UNet (UUNet) and Random Sampling-based UNet (RS-UNet) – align with the need of trustworthiness in the medical domain by involving domain experts in annotating datasets required for each iteration of the model. Furthermore, WD-UNet, LC-UNet, UUNet and RS-UNet achieve or even exceed the predictive level of some state-of-the-art DL models in the literature, while requiring less annotated training data (with only 15% - 35% of the training data). In conclusion, the aim of Deep Human-Computer Interaction is to enhance model performance while reducing the annotation cost for experts, thereby leveraging domain knowledge of radiologists on trachea and centerlines to ensure more accurate model predictions.

## 1.3 Human-Computer Interaction-based learning models

Figure 1 illustrates conceptually the challenges and contribution above stated. Through the revisions made by our medical experts, notable enhancements can be achieved in the accuracy of model predictions. Human-Computer Interaction-based learning models include LC_UNet, RS_UNet, UUNet and WD_UNet.

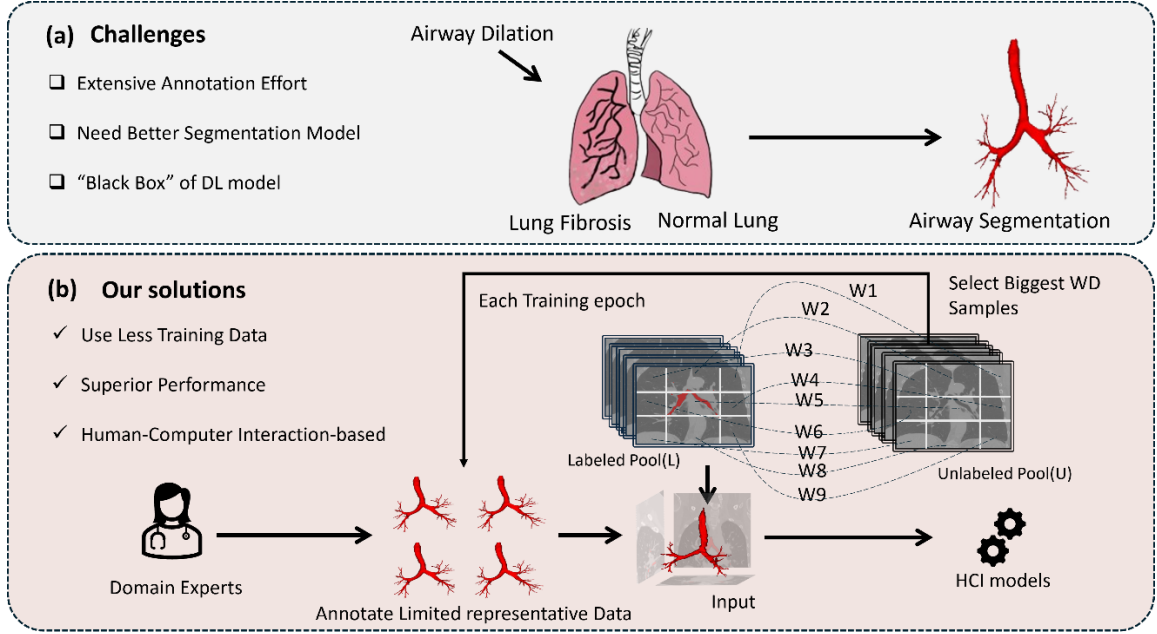

*Figure 1. Abnormal dilation of the airways caused by pulmonary fibrosis, current challenges, and our solutions. (a) Pulmonary fibrosis can lead to abnormal expansion of the airways, primarily characterized by thickening of the airways, dilation at the ends, and irregular shaping. (b) Domain experts only annotate part of the unlabeled data and update the training dataset for every epoch training of DL model. W means that the distance between pixels helps to measure the difference between two probability distributions.*

Deep learning (DL) has demonstrated remarkable efficacy across various medical image segmentation tasks, such as diagnosing lung diseases and delineating organ/cortex boundaries[11], [12]. While DL models hold the potential to reshape clinical practices, acquiring and annotating medical 3D HRCT image data is costly and labor-intensive, posing a major challenge in developing accurate deep learning models for medical image analysis. This challenge is particularly pronounced when dealing with extensive 3D medical segmentation datasets containing densely distributed elements. In contrast, human-computer interaction-based learning offers a strategic solution. Our Human-computer interaction-based learning models can choose the most uncertain unlabeled samples for manual annotation by domain experts (as depicted in Figure 2). This strategy is employed due to the recognition that the most uncertain samples harbor a wealth of informative features essential for training, thereby mitigating the laborious and costly nature of the annotation process[8]. Traditional active learning (AL) (also known as query learning) techniques are considered a specific retrieval strategy. They provide solutions by selecting and labeling a small group of the most informative and representative points from the pool of unlabeled data [13]. While Active Learning (AL) is also suitable for addressing the issue of high costs associated with acquiring labeled data, it falls short in handling the features of high-dimensional images in pixel-level segmentation and fails to extract complex spatial feature representations such as those of the trachea. Consequently, traditional AL is not applicable to the task of 3D airway segmentation[14].

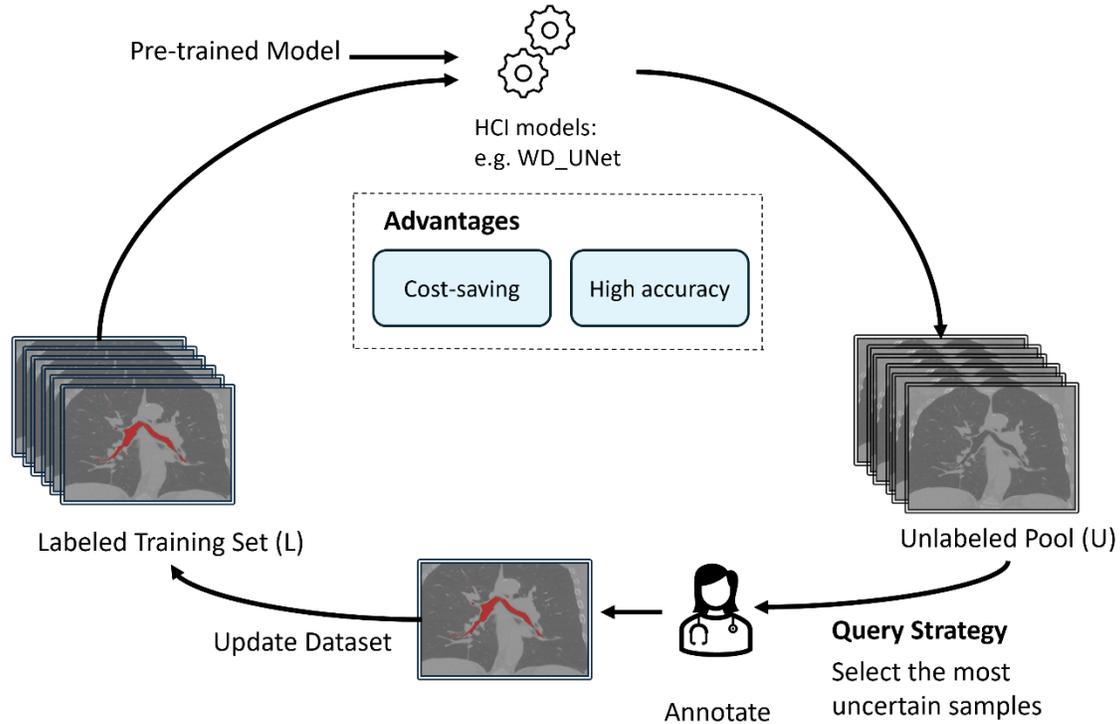

*Figure 2. Human-Computer Interaction-based deep learning querying cycle*[8]. * Labeled dataset (L) will be updated by labeling the selected samples from unlabeled pool in each Active Learning loop[15]. *The advantages refer to the strengths of our proposed HCI models compared to traditional Active Learning (AL) models.*

Human-Computer Interaction-based deep learning represents a fusion of human-computer interaction-based learning principles and deep learning models, aimed at pinpointing data samples deserving of annotation. As depicted in Figure 3, the approach unfolds as follows: (1) Initially, deep learning models undergo pre-training on a small, labeled dataset. (2) Subsequently, these models undergo iterative training, where query strategies based on uncertainty or least confidence guide the selection of the most informative or equivocal samples. (3) The chosen samples are then meticulously labeled by domain experts and integrated into the training dataset, prompting model retraining. (4) This cyclic process persists until the desired model performance is achieved or the annotation budget is fully utilized. Human-Computer Interaction-based deep learning harnesses the formidable information extraction capabilities and high-dimensional image processing prowess inherent in deep learning models. Meanwhile, the trachea segmentation DL model takes advantage of active learning's efficiency to curtail annotation expenditures[16]. In comparison to traditional low-shot segmentation approaches and co-segmentation methods, we proposed four Human-Computer Interaction-based deep learning models, for example, which distinguishes itself by its independence from pre-training quality, as well as its robust generalization and stability. Its independence from pre-training quality is primarily reflected in the fact that the final prediction of WD-UNet does not rely on the initial performance of a pre-trained model; instead, it improves through continuous iteration. The emphasis is on the training data updated in each iteration, highlighting the capacity for continuous learning (as new data can be incorporated easily into the model) and reducing dependence on large-scale data. Training data logs reveal that using only 15% to 35% of the training dataset, the evaluation score—primarily based on the Branch Detection Ratio—increased from 28% to 91%. This improvement occurred over 15 iterations of the query strategy, beginning with the data predicted by the pre-trained model and culminating in the training of the most effective model.

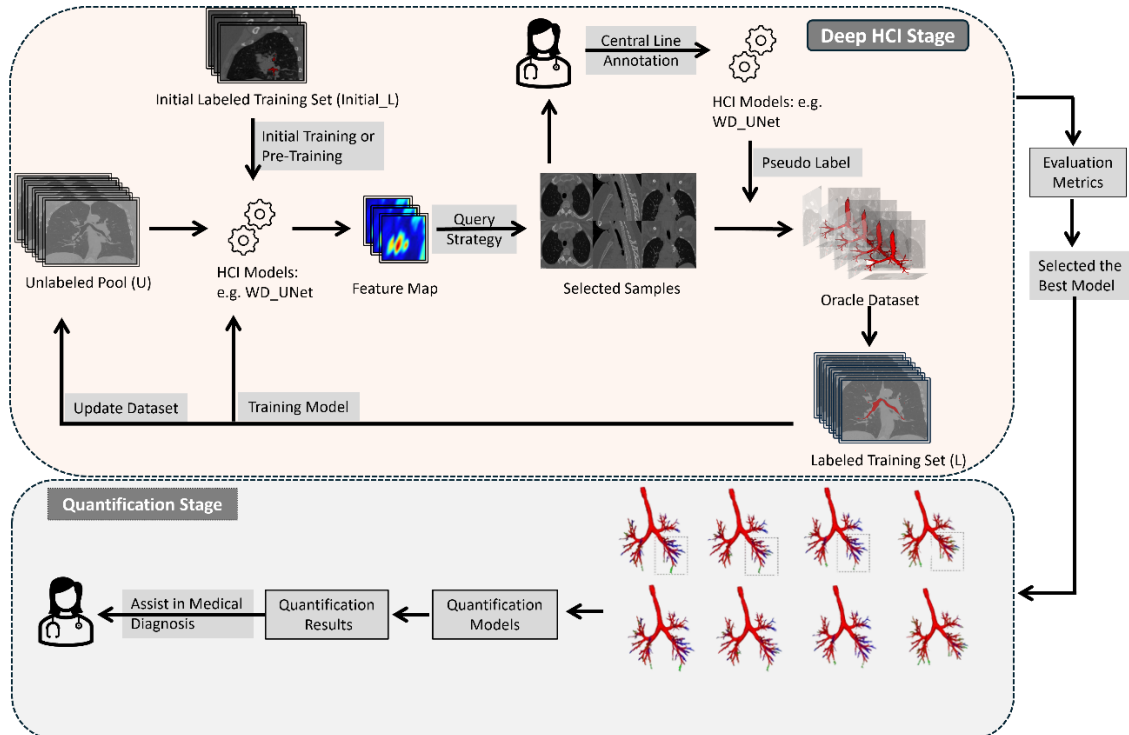

*Figure 3. Our framework outlines the structure of a typical Human-Computer Interaction-based deep learning. The process begins with the DL model defining parameters, often based on initial training with a small, labeled dataset (L0). The DL model then extracts features from the unlabeled data pool (U). A query strategy selects representative samples added to the labeled training dataset (L). This labeled dataset is used to update U and fine-tune the DL model in each iteration. This iterative process continues until predefined termination conditions and label budget are met. *L0 represents the initial training dataset, while L is the dataset updated after each active learning loop. After each loop, U's data is labeled and moved to L*[15]. *DeepHCIL, Human-Computer Interaction-based deep learning.*

**1.4 Findings**

Subsequently, we enlisted expert assistance to extract the tracheal central line as a corrective element added to the training process. Comparatively, we observed pronounced variations in quantitative results before and after the addition of the central line. The refinement by medical professionals resulted in more accurate central line segmentation, consequently leading to substantial improvements in the metrics used to evaluate predictive results, such as the Dice Similarity Coefficient (DSC), Intersection over Union (IoU), and precision. Simultaneously, the removal of erroneously segmented portions (labeled as false positives) led to slight decreases in branch score (BD) and tree detected ratio (TD). However, these findings underscore the clinical significance of incorporating the central line. This enhancement enhances the accuracy of the model's predictions, thereby offering better diagnostic support for patients with airway-related diseases, which is the primary objective of this study.

**1.5 Ethics**

The use of patient data in the Human-Computer Interaction-based learning models raise significant privacy concerns. Adhering to regulations like HIPAA and GDPR is crucial for protecting patient privacy. Implementing data anonymization and secure data handling protocols is essential. The datasets we used are all from public sources and comply with usage standards. Additionally, medical applications of Human-Computer Interaction-based learning models require rigorous testing and validation, as well as ethical and regulatory approvals. Meanwhile, this study was approved by the Ethics Committee on Biomedical Research of relevant hospitals, and the informed consent was waived (IRAS Project ID: 234527).

However, the Human-Computer Interaction-based learning models also have limitations, as its effectiveness highly depends on the quality and representativeness of the training data. Insufficient or biased data can significantly impair the model's performance and generalizability. Therefore, our selected samples include a variety of ages, genders, normal/with severe illness/lung cancer, etc., to ensure dataset diversity.

For all the sections of this paper, a brief summary can be described as:
1. Introduction encompasses the challenges of HRCT airway segmentation (Section 1.1), the contributions of our proposed HCI models—RS_UNet, LC_UNet, UUNet, and WD_UNet (Section 1.2), a brief overview of the workflow and principles of the proposed HCI models (Section 1.3), related findings of the article (Section 1.4), and considerations of ethics (Section 1.5).
2. Related Work section includes introductions and comparisons between supervised models (Section 2.1) and our proposed HCI models (Section 2.2), also referred to as semi-supervised models.
3. The Methodology section details the dataset information and division (Section 3.1), the supervised model benchmarks we use for comparison (Section 3.2), a focused description of one HCI model—WD_UNet (Section 3.3), and descriptions of the other three HCI models, LC_UNet, RS_UNet and UUNet (Section 3.4).
4. The Results and Discussion section includes an introduction to the experimental setup and equipment (Section 4.1), qualitative and quantitative results analysis (Section 4.2), and visual prediction analysis of different models for airway segmentation (Section 4.3).
5. Conclusion substantiates that the proposed HCI models can achieve or even surpass the performance of supervised models while only utilizing 15%-35% of the training data, thereby reducing the annotation effort.

## 2      Related Work

The primary challenges we seek to overcome are to enhance the authenticity of tracheal segmentation to achieve more accurate diagnoses for patients suffering from lung diseases. By doing so, we not only prioritize the health and well-being of our patients, but also aim to reduce the burden of labelling costs on radiologists, making their work more efficient and effective. We briefly list several approaches that build deep learning models for medical image segmentation in the following. The approaches are divided into two categories: Supervised DL model and semi-supervised DL model. Supervised deep learning models rely solely on annotated data, while semi-supervised models use both annotated and unannotated data for the modelling task at hand. Supervised models depend on data quality and quantity, whereas semi-supervised models are more flexible with limited annotations.

**2.1 Supervised DL Models**

In recent years, 3D UNet has gained prominence as a rapidly evolving and popular category of segmentation algorithms designed for 3D image processing. UNet is characterized by its contracting path/up-sampling for capturing context and symmetric expanding path/skip connections for precise localization, retaining local information while capturing global context. This aids in restoring the fine details required in medical imaging. UNet employs multi-level convolution and pooling operations to efficiently extract features at different levels, making it suitable for multi-scale and complex medical images. Another key reason we chose 3D UNet as the backbone for our proposed Human-Computer Interaction-based deep learning models is its efficient training process and lower number of parameters.

This allows UNet to maintain efficient training and good performance even with limited computational resources. The exceptional performance of 3D UNet has garnered considerable attention and adoption among users, leading to the development of various adaptations, such as UNet++[17], nnUNet[18], VB-Net[19], Fully Convolutional Network (FCN)[20], SegNet[21], etc. have demonstrated great performance of medical image segmentation. UNet++ introduces nested, dense skip connections into the original UNet architecture, enhancing the model's ability to capture details in boundary areas. nnUNet is an adaptive framework that adjusts the network's depth, width, and other hyperparameters based on the characteristics of the data. VB-Net improves recognition of vascular structures through more complex feature extraction and representation learning methods. FCN, like UNet, incorporates the concept of upsampling and skip connections. SegNet has a lighter structure compared to UNet, but it may not capture details as finely as UNet. Furthermore, tools like the MONAI Framework, a PyTorch-based open-source AI framework jointly developed by NVIDIA and King's College London in late 2019, provide a free and community-supported platform for deep learning in medical imaging segmentation[22][23]. The MONAI framework offers various variants of the original UNet architecture, optimized for three-dimensional medical imaging. However, many of these models are essentially variants of the vanilla 3D UNet. In the context of specialized tasks, such as lung airway segmentation, they may not provide the best predictive results tailored to the unique characteristics of the airway. Like other UNet variant networks, MONAI also is not specifically designed for airway segmentation tasks. For instance, typical loss functions in medical segmentation, like Dice loss[24] and BCE loss[25], focus on general features. To specifically learn airway features, we introduce Branch loss, targeting the airway's structure. We also apply a Central Loss Function, using machine-extracted central lines refined by experts as ground truth in training. This approach improves IoU, Precision, and Dice score of all these supervised models and proposed HCI models. The difference between these models will be discussed in Section 3.2-3.4, while their performance analyzed in Section 4.2.

**2.2 Semi-supervised Model**

Semi-supervised learning models offer various advantages, including the more efficient utilization of data, reduced annotation costs, improved generalization performance, and applicability to scenarios with scarce labeled data[26]. The proposed four HCI models are all considered as semi-supervised models and the training of HCI models explicitly requires the intervention of experts at every query round.

**Query Strategy.** In the realm of Human-Computer Interaction-based learning models, query strategies play a pivotal role in enhancing the efficiency and effectiveness of model training. These strategies primarily bolster the annotation process by judiciously selecting the most informative samples labelled by domain experts, thereby optimizing the use of limited annotation resources. Such a targeted selection approach accelerates the learning process of the model, particularly around the complex decision boundaries, leading to quicker performance improvements. Furthermore, by reducing the quantity of samples required to achieve comparable performance, query strategies effectively lower the overall annotation costs. Another significant advantage is the enhancement of the model's generalizability; by learning from a more diverse set of sample features, the model becomes more robust to varying inputs. However, a notable drawback is the increased computational cost, especially with strategies based on model uncertainty, which require additional computational resources for assessing the uncertainty levels of samples. Deep human-computer interaction-based learning model is one kind of the semi-supervised models. When selecting unlabeled samples through query strategies, there are various strategies to choose from. These strategies include methods like Least Confidence[7], Margin Sampling[27], Entropy Sampling[8], Cluster-Based Selection[28], Bayesian Active Learning Disagreement[29], and many more. Least Confidence picks the most uncertain samples for single-label classification. Margin Sampling selects samples with the smallest probability difference between the top two classes, focusing on marginal decisions. Entropy Sampling chooses samples with high information entropy, indicating uncertainty across all classes. Cluster-Based Selection identifies representative samples from clusters in diverse datasets. Bayesian Active Learning Disagreement (BALD) selects samples based on model prediction variances, considering the model's inherent uncertainty. The Wasserstein Adversarial strategy[24] employs the Wasserstein distance from GANs to assess the potential value of unlabelled samples for model training. Its goal is to select samples that most effectively aid the model's learning and narrow the gap between the true data distribution and the model's predictive distribution. In our paper, since Least Confidence is suitable for single classification (for example, Cat/Dog, True/False), it can also be applied to tracheal segmentation with binary labels (0/1). Entropy Sampling predicts the entropy of each pixel, where high entropy indicates high uncertainty, and low entropy indicates low uncertainty. Additionally, we proposed another query strategy, Random Sampling, which randomly selects n unannotated samples. Overall, we combine these three query strategies with 3D UNet, along with WD-UNet, to form our proposed four Human-Computer Interaction-based deep learning models. WD-UNet is the best performing model among all the benchmark models and other three proposed semi-supervised models, we now describe it in further detail.

Similar Human-Computer Interaction-based learning work has been implemented in various contexts. For example, in Huang K's article "DeepAL: Deep Active Learning in Python"[30], four public datasets - "MNIST," "FashionMNIST," "SVHN," and "CIFAR10" - were used, and several query strategies were applied like Least Confidence, Entropy Sampling, Random Sampling. Along with 2D network structures, all related to image classification of everyday objects or animals, but not specifically to medical segmentation. Some surveys of Deep Active Learning mentioned about few methodologies of medical image segmentation, but not specially to airway segmentation[31][32]. Therefore, Human-Computer Interaction-based learning approaches in the context of medical segmentation, like the ones proposed in Huang K's article[30], have not been explored. Drawing inspiration from the paperwork of Huang K[30], we have developed Human-Computer Interaction-based learning segmentation models tailored to the characteristics of pulmonary airways, including LC-UNet, UUNet, RS-UNet, and WD-UNet. Furthermore, our proposed annotation framework using deep learning can be integrated with various query strategies and any 3D segmentation network structure.

The remaining common semi-supervised models for image segmentation include Mean-Teacher[33], which requires training two models and hence implies more computational resources. Particularly in environments with limited resources, this requirement can become a significant obstacle for practical applications. For scenarios seeking high efficiency and cost-effectiveness, Mean-Teacher may not be the optimal choice. FixMatch[34], is another semi-supervised image segmentation approach, which enhances model performance by combining weak and strong data augmentation strategies. However, the effectiveness of FixMatch heavily relies on the appropriate selection and implementation of data augmentation strategies. Its application may be limited in scenarios where data augmentation strategies are unclear or challenging to execute. Moreover, due to the complexity of its algorithm, implementing and optimizing FixMatch could be difficult for beginners or researchers with limited computational resources. Self-training[35] is a simpler alternative but may perform poorly when dealing with low-quality unlabeled data. This method relies on the model's initial predictions for the unlabeled data, and inaccurate predictions can lead to error accumulation and, subsequently, impact the final model performance. Therefore, for tasks with varying quality of unlabeled data, self-training might not be the best solution. Pseudo-Labeling[36] is also a semi-supervised option for image segmentation, yet it demands higher model performance and pseudo-label quality, making it less suitable for all tasks. It uses model predictions as pseudo-labels for further training. However, if the initial model performance is poor, errors in pseudo-labels can misguide the model's learning path. Thus, in cases with weak initial performance or low-quality pseudo-labels, Pseudo-Labeling might not be ideal. However, our proposed Human-Computer Interaction-based learning model has not yet been applied to the segmentation of lung airways. It aims to circumvent the limitations associated with other semi-supervised models while also offering the advantage of requiring low computational resource (GPU usage).

# 3     Methodology

## 3.1 Dataset

We utilized a dataset comprising 140 cases sourced from a combination of the EXACT09 dataset[37] and the LIDC-IDRI dataset[38], featuring 3D HRCT data. The dataset was partitioned into training, validation, and test/inference sets, comprising 72, 18, and 50 cases, respectively, with each case containing original images and corresponding ground truth. The first step in data pre-processing involves splitting the data. The test dataset remains unchanged. The ratio between training and validation must remain constant, but in different iterations of cross-validation, the validation dataset needs to consist of completely independent and non-overlapping patches. The patches for the validation set have been pre-split, constructing two completely different train/valid groups, with the same ratio but different patches.

*TABLE I The labeling and proportion of our dataset.*

| Dataset | EXACT09 + LIDC-IDRI | | |
|---|---|---|---|
| | Labeled (patches) | Unlabeled(patches) | Total(patches) |
| **Train** | 356 | 660 | 1016 |
| **Validation** | 261 | N/A | 261 |
| **Test** | 50 CTs | N/A | 50 CTs |

It merits emphasis that our training and validation datasets are distinctly segregated from the test dataset, encompassing entirely disparate sets of data. Initially, a cross-validation method employed during the training period ensures continuous assessment of the model's performance against unseen data. Furthermore, the training and validation datasets are composed of 3D CT scans of normal pulmonary structures, accompanied by corresponding ground truths. Conversely, the test dataset is derived from 3D CT pulmonary tracheal segmentation datasets of patients afflicted with COVID-19 and pulmonary fibrosis. By distinctly separating normal pulmonary structures in the training and validation datasets from pathological cases in the test dataset, the model is challenged to generalize well to scenarios it has not encountered before. These strategies collectively enhance the model's generalization capability towards new, unseen data, effectively mitigating the risk of overfitting despite the limited size of the labeled training dataset.  Due to the impracticality of utilizing complete 3D CT images as input for deep convolutional networks (DNN) due to their size, preprocessing was performed by segmenting the data into 3D patches of dimensions [128, 96, 144], each containing parts of the pulmonary airway region. The varying number of patches per case is due to the differing thicknesses of each case, i.e., the Z-axis. This preprocessing step generated a total of 1277 patches. Among these, 356 patches were allocated for initial training, while the other 660 patches remained unlabelled. An additional 261 patches were set aside for accuracy assessment following each iteration of the DeepAL method's querying cycle. The remaining 50 cases were exclusively designated for inference, ensuring no overlap with the training and validation datasets.

## 3.2 Supervised Learning Benchmarks

To conduct a comprehensive comparative analysis involving semi-supervised Interaction-based Learning approaches, we have introduced two supervised baseline models: the 3D UNet[23], [39] and the CEUNet[6]. This introduction serves the purpose of facilitating a performance comparison between our proposed HCI algorithms (LC-UNet, UUNet, RS-UNet, and WD-UNet) and these supervised models. In our forthcoming research, we present an adaptation of the UNet architecture known as CEUNet. CEUNet incorporates a Dense Atrous Convolution (DAC) module and a Residual Multinuclear Pooling (RMP) module within the architecture, strategically positioned between the encoder and decoder components[40]. This architectural enhancement empowers CEUNet to employ convolutions with varying dilation rates, thus enhancing its ability to extract features from objects of diverse sizes and generating feature maps with variable resolutions. It is worth mentioning that the idea is derived from Zaiwang's concept[40]. In his paper, comparisons with experimental results involving UNet are also made. However, his approach is solely based on 2D images and has never been applied to 3D contexts. Inspired by his work, we have developed the corresponding CEUNet benchmark to conduct comparisons alongside UNet.

### 3.3 Semi-supervised (Human-Computer Interaction-based deep learning) novel method: Proposed WD-UNet

As the algorithms employed in the proposed Wasserstein Distance-based UNet, such as the query strategy, have never been utilized before, this section is dedicated to a detailed description of the main principles underlying the construction of WD-UNet.

The proposed WD-UNet model is built upon the foundation of the 3D UNet deep learning (DL) model. In each human-computer interaction-based learning querying cycle, this DL model undergoes a retraining process guided by query strategies aimed at identifying the most informative data for annotation. This iterative approach ensures that the model is continuously updated until it achieves the desired performance. The training procedure, illustrated in Figure 4, consists of two main steps:

- The initial labeled data is fed into the model for training, during which the Dice, BCE, and Branch Loss are computed.
- Both the labeled and unlabeled data are sequentially passed through the feature extractor and the Discriminator.
    1) In the context of model training, the utilization of a "discriminator" proves valuable in the assessment of distinctions between labeled and unlabeled data, with its outcomes being incorporated as one of the components contributing to the loss function. In scenarios where unlabeled data encompass instances exhibiting inherent ambiguity, marginal characteristics, or potential mislabeling, the employment of a "discriminator" may facilitate the discernment of those unlabeled samples possessing genuine informative value. This can, consequently, lead to a considerable improvement in the overall performance of the model. This process yields

---

**Algorithm 1** Gradient Penalty Approach

**Require:** Discriminator network, unlabeled data(h_s), labeled data(h_t)

**Ensure:** Penalty value

1: alpha = torch.rand ()

2: interpolates = h_s + (alpha * (h_t – h_s))

3: preds = Discriminator(interpolates)

4: gradients = torch.autograd.grad(preds, interpolates)

5: **if** gradients.norm(gradients) > max_norm:

   clip_coef = max_norm / gradients.norm(gradients)

   gradients = gradients * clip_coef.

6: **end if**

7: penalty = ((gradients.norm(gradients)- 1) ** 2).mean()

8: **returns** penalty

extracted labeled features (lb_out) and extracted unlabeled features (ulb_out) (see Figure 4), which are then employed to calculate the Wasserstein distance between them.

2) Subsequently, lb_out, ulb_out, and the Discriminator are collectively processed through the Gradient Penalty function to determine the penalty. The penalty is computed by assessing the mean of the squared difference between the gradient norm and 1. Finally, the penalty is added to the Wasserstein distance, resulting in the Wasserstein Discriminator Loss (WD Loss).

3) These five loss functions are integrated to optimize the WD-UNet model, which contains WD loss, Dice loss, BCE loss, Central line loss and Branch loss.

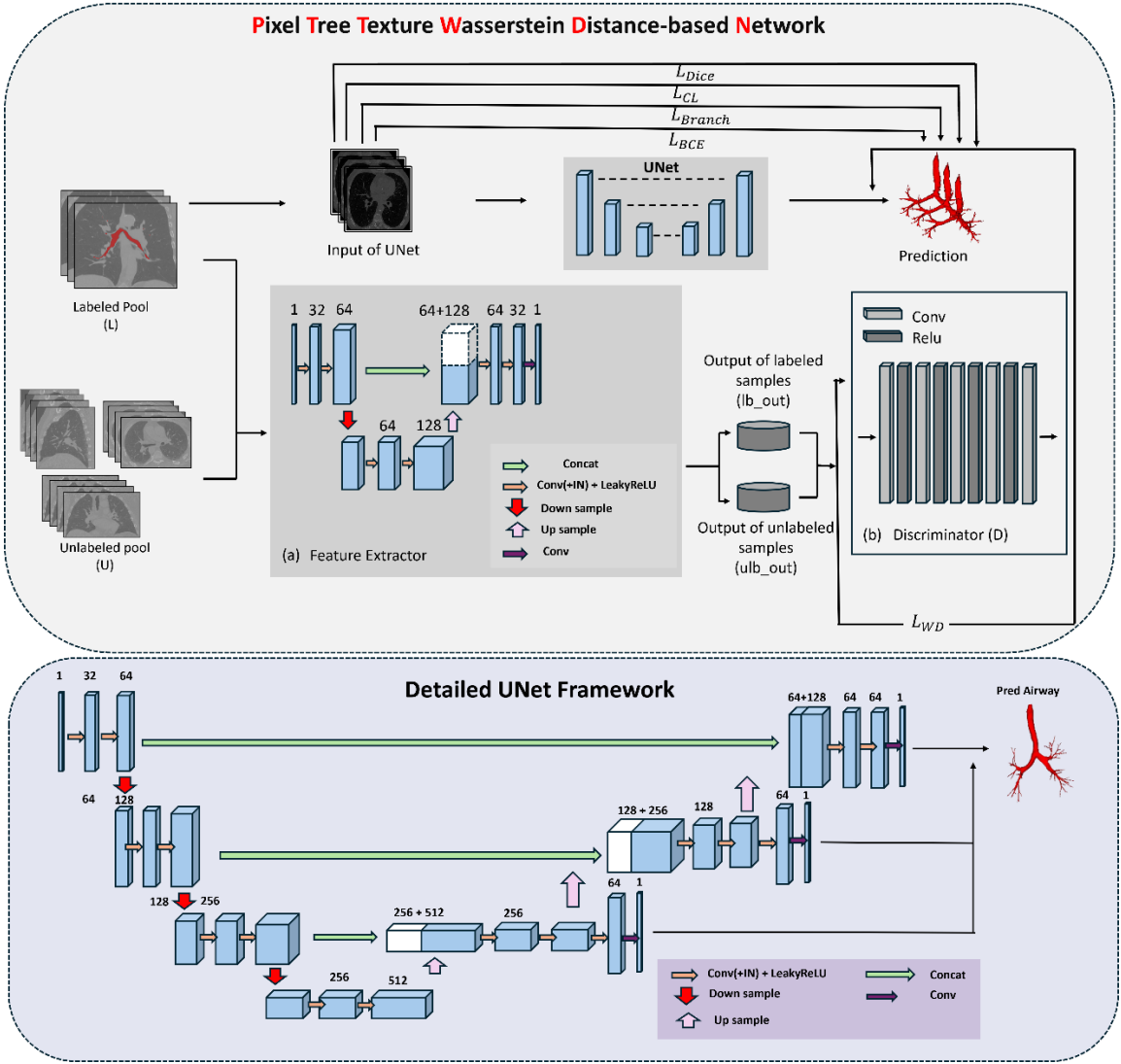

*Figure 4. The whole training procedure of WD-UNet. * A detailed explanation of Branch Loss, Discriminator, Feature Extractor, Wasserstein Distance, and Gradient Penalty can be found in the Methodology section. The comprehensive loss should incorporate Dice, BCE, Branch, Central Line and WD Loss components. Figure Description: (a) Provides a detailed overview of the Feature Extractor, which can be conceptualized as a simplified UNet structure, including one downsample, one upsample, and a skip connection. Its primary purpose is the efficient extraction of feature representations from both labeled and unlabeled images. (b) Illustrates the architecture of the Discriminator, consisting of five Convolution layers and four Rectified Linear Units (ReLU) activations. * The annotated and unannotated data are not paired in the initial dataset. "lb_out" means output of labeled samples, "ulb_out" means output of unlabeled samples. "$L_{Dice}$", "$L_{CL}$", "$L_{Branch}$", "$L_{BCE}$", "$L_{WD}$" are loss functions.*

**Wasserstein distance,** also referred to as Earth Mover's Distance (EMD)[41], [42], Wasserstein distance proves to be a valuable metric for quantifying dissimilarity between two distributions, especially in tasks related to image generation and transformation. When compared to alternative distance measures such as Kullback-Leibler (KL) divergence[43] or Jensen-Shannon (JS) divergence[44] Wasserstein distance excels in precisely assessing differences between distributions[45], even in intricate scenarios involving mode collapse, while providing superior gradient information. The theoretical foundation of Wasserstein distance lies in Optimal Transport theory[46] cost necessary to transform one distribution into another. In our proposed WD-UNet model, Wasserstein distance is uniquely harnessed to distinguish between unlabeled and labeled empirical distributions, where their dissimilarity contributes to the loss functions for WD-UNet. This mechanism facilitates the selection of diverse training data. The computation of unlabeled and labeled empirical distributions is executed through a dedicated discriminator. Concerning uncertainty, the primary goal is to identify samples with lower predictive confidence, indicative of higher uncertainty. In terms of diversity, the objective is to pinpoint unlabeled batches incurring greater transportation costs under Wasserstein distance when contrasted with the labeled set, a reliable indicator of dissimilarity from the existing labeled samples, thus enhancing the model's diversity.

**Gradient penalty.** The function used to compute the gradient penalty for training the discriminator network and calculating the Wasserstein distance plays a crucial role in our approach. This function receives inputs such as the discriminator network, unlabeled data(h_s), and labeled data(h_t). The primary purpose of the gradient penalty is to incentivize the discriminator to maintain gradients close to 1 for interpolated samples. This adjustment greatly enhances the model's training effectiveness and stability. Furthermore, this module is responsible for calculating the penalty, which represents the mean of the squared difference between the gradient norm and 1. This penalty value serves a dual purpose as a gradient penalty and as one of the loss functions for model updates.

**Feature extractor.** The feature extraction module is an essential component designed to process raw data efficiently. It serves the primary function of converting raw data into a discriminative representation, which is pivotal for enabling active learning training. By applying feature extraction to both unlabeled and labeled data, the module achieves the following outcomes:

- It generates lower-dimensional feature representations that capture essential information within the data. These representations aid in better comprehension and differentiation of various classes or attributes, particularly when calculating the Wasserstein distance in the Discriminator.
- The module filters out irrelevant information, enhancing the model's efficiency and its ability to generalize effectively. This step contributes to improved model performance.

**Wasserstein Discriminator Loss ($L_{WD}$).** The WD Loss (Wasserstein Discriminator Loss) is a composite of Wasserstein Distance and Gradient Penalty (GP). It is responsible for quantifying the diversity or dissimilarity between unlabeled data and labeled data. This process results in higher transportation costs under the Wasserstein distance metric and enhanced gradient information to ensure model stability and diversity[15]. In Equation 1, $lb_{out}$ and $ulb_{out}$ represent the outputs from the feature extractor for labeled and unlabeled data samples, respectively. The term $L_{Wasserstein\ Distance}$ denotes the dissimilarity between two feature distributions as calculated by the Wasserstein Discriminator, which is trained to distinguish the most dissimilar samples. Additionally, incorporating a penalty value as part of the loss function enhances the stability of the Wasserstein Discriminator and mitigates the risk of overfitting.

$$L_{Wasserstein\ Distance} = D(ulb_{out}) - D(lb_{out}).$$

$$L_{Penalty} = GP(D,\ ulb_{out},\ lb_{out}).$$

$$L_{WD} = L_{Wasserstein\ Distance} + L_{Penalty}. \quad \text{(Equation 1)}$$

**Branch Loss ($L_{Branch}$).** The term "index" denotes the index assigned to each branch, and each branch is determined through a calculation based on the parent-children relationship[47]. The process involves leveraging tree parsing techniques to compute the branches associated with each segment. The intersection value is acquired by taking the product of the predicted branch for each segment and the corresponding ground truth (GT) branch, followed by summing up these products. The denominator is calculated by summing the branches present in the ground truth for each segment. Consequently, the branch loss can be mathematically expressed as follows:

$$L_{Branch} = 1 - \frac{\sum Pred_{index} * GT + smooth}{\sum GT_{index} + smooth}. \quad \text{(Equation 2)}$$

In the context of WD-UNet, a specialized loss function has been formulated to harness the unique attributes of lung tracheal segmentation for optimizing the segmentation performance of a deep learning model. This loss function significantly enhances segmentation accuracy by focusing on the intricacies of the model's detail-to-structure performance. Unlike conventional deep learning algorithms for medical segmentation, which primarily rely on Dice Loss (dice similarity coefficient) and BCE (binary cross entropy) Loss weights, the newly proposed training loss function capitalizes on the distinctive characteristics of the trachea. It computes the Branch Loss of tracheal branches, which quantifies the degree of alignment between the predicted airway centerline and the ground truth airway centerline. A Branch Loss value of 0 signifies a perfect match, while a value of 1 indicates no overlap between the two centerlines. To ensure numerical stability, a smoothing factor is introduced to prevent division by zero.

$L_{Dice}^{w}$ represents Dice Loss, $L_{BCE}^{w}$ represents BCE Loss. The overall loss function can be represented as Equation 3:

$$L_{total} = L_{Dice}^{w} + L_{BCE}^{w} + L_{branch} + L_{WD}. \quad \text{(Equation 3)}$$

**Central Line Loss ($L_{CL}$).** The calculation method for central line loss is similar to branch loss, with the additional step of extracting skeletons from both the predicted branch and the ground truth. This is achieved by employing image erosion to extract the complete 3D airway's central line, where the airway's internal diameter is defined as a single-pixel size.

In the context of the erosion() function, it signifies the process of airway erosion, resulting in the extraction of a one-pixel-wide tracheal skeleton, commonly referred to as the central line. Consequently, the central line loss can be expressed as follows:

$$E_{pred} = f(x) = erosion(Pred_{index}).$$

$$E_{GT} = f(x) = erosion(GT_{index}).$$

$$L_{CL} = 1 - \frac{\sum E_{pred} * E_{GT} + smooth}{\sum E_{GT} + smooth}. \quad \text{(Equation 4)}$$

Therefore, by incorporating the Dice Loss (dice similarity coefficient) and BCE Loss (binary cross entropy), the overall Loss function can be obtained:

$$L_{total} = L_{Dice} * w_1 + L_{BCE} * w_2 + L_{Branch} * w_3 + L_{CL} * w_4 + L_{WD} * w_5. \quad \text{(Equation 5)}$$

In the context mentioned, w1, w2, w3, and w4 represent the weights multiplied with different loss functions. These weights can be tuned according to individual requirements and preferences, e.g. [w1,w2,w3,w4,w5]=[0.2,0.2,0.2,0.2,0.2]. The total loss measures how well the predicted airway matches the GT, with a value of 0 indicating a perfect match and a value of 1 indicating no overlap between the two. The smooth factor is added to avoid numerical instability caused by division by zero.

**Central Line Extraction.** The algorithms for branch loss and central line loss mentioned earlier were employed with the airway centerline extracted using the Python package skeletonize_3d() in the absence of medical expert correction. To investigate the impact of using expert-corrected centerlines on the experimental results, a comparative study was conducted (In Table III). Consequently, an additional set of experiments was established in which, during each round of Human-Computer Interaction-based deep learning query strategy selection, a radiologist corrected the preliminary airway centerlines extracted, and this corrected version was used as a training loss for the model. This training loss can be represented as:

$$L_{CL} = 1 - \frac{\sum \text{Pred}_{CL} * \text{GT}_{CL} + smooth}{\sum GT_{CL\_} + smooth}. \qquad \text{(Equation 6)}$$

**Query Strategy.** The process begins with the extraction of indices and data associated with the unlabeled samples. Subsequently, the probability prediction method is employed to generate probability predictions for the unlabeled data, yielding a set of probability values. Two distinct scores are then computed:

- The uncertainty score, determined by giving weight to the upper bounds of both the L2 norm and the L1 norm.

- The discriminative score, derived from the predicted output of the unlabeled data using the Discriminator.

Finally, the total score is obtained by subtracting the product of the discriminative score from the product of the uncertainty score, taking into consideration the selection parameter.

**Human-AI collaboration.** Using our Human-Computer Interaction-based deep learning framework, various query strategies and deep learning models can be nested. Taking WD-UNet as an example, when working with only raw lung HRCT data (all unlabeled), the workflow, also shows in Figure 3, is as follows:

- **Initial training.** Radiologists initially annotate a limited subset of HRCT lung data, from which the machine autonomously generates airway central lines. These central lines are subsequently refined by experts, creating the initial dataset comprising original CT scans, ground truth labels, and central lines.

- **Query strategy for selecting unlabeled data.** Through an initial training model (Round 0), predictions are generated from the unlabeled dataset. Employing L1 and L2 norms, the most uncertain and information-rich samples are selected as part of the unlabeled data selection strategy. Ten of these samples (the number can be customized for each expert annotation round) are then designated for expert annotation.

- **Human Collaboration**. After experts annotate the location of airways, machine-generated airway central lines are refined by the experts. Consequently, the Human-Computer Interaction-based deep learning models obtain ten newly annotated cases (including airways and central lines) in each round. These newly annotated cases are then incorporated into the labeled training dataset.

- **Iterations.** Following a customized number of Human-Computer Interaction-based deep learning training iterations (includes query strategies -> expert labeling -> central line manual corrections -> training with new data -> query strategies), the final Human-Computer Interaction-based deep learning model is attained.

**Post-Processing.** Maximum connectivity calculation: Connected Component Analysis (CCA) is used as a common post-processing technique in a number of state-of-the art algorithms [48]. This technique is used to isolate individual components of the segmentation output using connected neighborhoods and label propagation. In order to filter out unwanted and disconnected segments under a threshold, the skimage.measure.label() function can label connected regions of a binary image. The return value of the function is the labelled image and the number of assigned labels, and does not change the binary image. A NumPy array is created to store the volume of each labeled region in the input image. A loop iterates through each labeled region, and uses NumPy's boolean indexing and sum() function to calculate the volume of the region. Then the volume array is sorted in ascending order and the largest region is selected and kept by indexing.

### 3.4 Semi-supervised (Human-Computer Interaction-based deep learning) novel methods: Proposed LC-UNet, UUNet, RS-UNet

Among the three proposed Human-Computer Interaction-based deep learning models, namely Random Sampling UNet (RS-UNet), Least Confidence UNet (LC-UNet), and Uncertainty UNet (UUNet), it's important to note that the latter two employ entropy-based principles for sample selection. Among the four HCI approaches we proposed, both least confidence and random sampling have previously been implemented as query strategies in traditional active learning. The query strategy employed by UUNet is inspired by the least confidence algorithm, wherein the entropy for each pixel is calculated based on its probability value using the formula $-p * \log_2(p)$. The entropies of individual pixels are then accumulated, and the average entropy is used as a metric for modeling uncertainty. Thus, the strategy mentioned in UUNet represents a novel query strategy.

The proposed LC-UNet, UUNet, and RS-UNet differ from the proposed WD-UNet in the following methods:

- **Query Strategy.** The query strategies of WD-UNet and UUNet are novel and have been newly proposed, distinguishing them from the traditional approaches employed by LC-UNet and RS-UNet.
- **Network Structure.** LC-UNet, UUNet, and RS-UNet utilize the classic UNet architecture, which contrasts with the specially designed, more complex network structures such as the Feature Extractor and Discriminator mentioned in WD-UNet.
- **Loss Function.** WD-UNet employs additional loss functions, the Wasserstein distance, combined with Gradient Penalty to train the Feature Extractor and Discriminator.

The proposed LC-UNet, UUNet, and RS-UNet share the following methods with the proposed WD-UNet: pre-processing, post-processing, branch loss, central line loss, and central line extraction.

Through extensive experimentation, it was found that WD-UNet outperforms both the two supervised models and the four proposed HCI models in terms of performance (see Table III). Consequently, WD-UNet is detailed in the methodology section.

## 4      Results and Discussion

### 4.1     Experiment Setup

**Coding Environment and Radiologist Annotation tools.** The experiments were conducted in an environment equipped with four GPUs, including a high-performance graphics card (specifically, four Nvidia GeForce 3090Ti with 24GB each). Various annotation tools, such as 3D Slicer[49], ITK-Snap[50], MeshLab[51], and MIMICS[52], were employed in the process. ITK-Snap is primarily used by domain experts for annotating pulmonary airways and correcting central lines. MeshLab is utilized to convert the NIFTI (.nii) format airway ground truth annotated in ITK-Snap into a mesh. 3D Slicer is employed for converting the 3D CT scans corresponding to the airway ground truth into the DICOM medical imaging format. Finally, the DICOM format CT and airway mesh are inputted into MIMICS to generate the central line, which is then annotated by experts. It's important to note that we are committed to sharing our relevant work and source code with the research community in the future. The models in all experiment tables were each run for 15 epochs, selecting 10 patches with the most informative information content for expert annotation each time, totaling 150 patches improved through annotation. 15%, 35%, 55%, and 75% training data represent using n=152, n=356, n=559, and n=762 patches, respectively, relative to 100% training data which consists of 1016 patches. Here, n is calculated as the initial training data numbers + 10*15. The exception is the 15% training data, which starts with an initial training data number of 50, runs for 10 epochs, and selects 10 patches each time.

**Evaluation Metrics.** Predictions were made on our test set and the results were compared with the ground truth annotated by the radiologist to calculate four metrics to measure the performance of the model, namely Dice similarity coefficient (DSC), Precision, Tree Dsetected ratio (TD) and Branch Detected ratio (BD). To statistically evaluate the performance, Wilcoxon signed-rank test was adopted between the evaluation metrics derived using DCGN and other comparison methods, with $P < 0.05$ (or $P < 0.001$) indicating significant (or highly significant) differences between the two paired methods.

**Correction of the lungs' airway central lines.** The "skeletonize_3d()" function from the skimage.morphology package, which is part of the external SciKit-Image Python library, is a widely adopted algorithm for extracting object skeletons from 3D images. However, since this function is designed for general 3D images and does not tailor its operation to the specific characteristics of the trachea, it may not yield smooth and clearly branched tracheal centerlines. Consequently, expert intervention is required to manually rectify the centerlines extracted through the "skeletonize_3d()" process, as depicted in Figure 5.

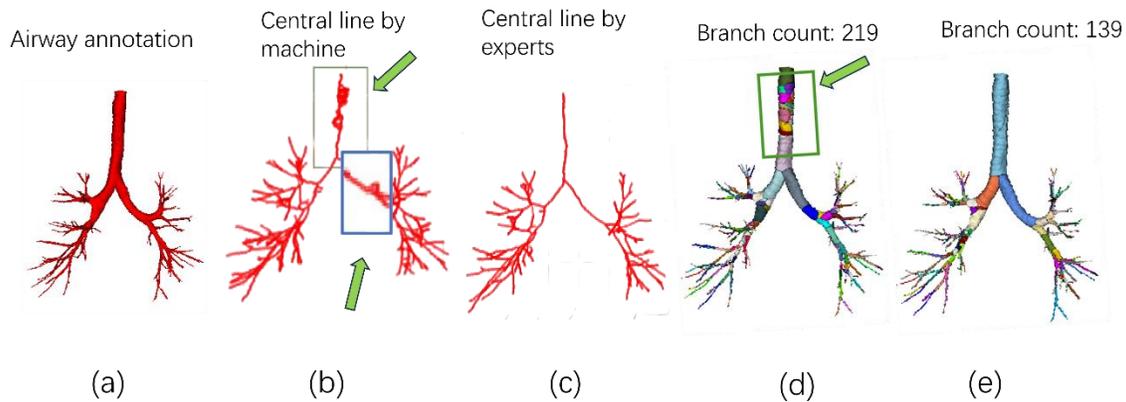

*Figure 5. Visualisation of airway parsing results. Central line extraction results by machine (b) and the revised central line by human radiologists(c). Branch-level labeling results based on machine extraction (d) and ours (e). Different branches were illustrated by different color.*

Upon observing Figure 5, it becomes evident that the centerlines extracted by the Python function exhibit a circular shape at both the main trachea (depicted in the green box in Figure 5(b)) and the smaller bronchi (highlighted by the blue box in Figure 5(b)). This characteristic is also reflected in the tree parsing results in Figure 5(d) and Figure 5(e). Inaccurate machine-extracted centerlines can lead to imprecise tracheal segmentation, thus affecting the accuracy of the final prediction results.

**4.2 Qualitative and Quantitative Findings**

Accordingly, 2-fold cross-validation is performed on the validation set. Each HCI model is considered a semi-supervised model, owing to the integral role of the human-in-the-loop approach within the training process, which is updated with each iteration of the query strategy round. Table II and Figure 6 reveal an important observation – WD-UNet, utilizing Wasserstein Distance and offering the advantage of lower computational complexity, stands out as the preferred choice. This model is subjected to training using varying proportions of the available training data (15%, 35%, 55%, and 75%, n= 152, 356, 559, 762 patches). The results demonstrate that WD-UNet excels when trained with 35% of the training data (n=356), exhibiting performance on par with the two supervised models, UNet and CEUNet. The 35% WD-UNet model is marginally lower than UNet in terms of the BD metric by a mere 0.08, achieving a score of 0.868. However, it surpasses UNet in all other evaluated metrics, namely DSC, Precision, TD, and IoU, with respective scores of 0.930, 0.895, 0.897, and 0.871. Furthermore, 35% WD-UNet outperforms CEUNet across all evaluation metrics. Notably, even the WD-UNet model trained with merely 15% of the training data (n=152) demonstrates exceptional predictive performance. While it underperforms compared to the 35% WD-UNet in terms of DSC, precision, IoU, and BD, it surpasses the latter in the TD metric. Additionally, it is noteworthy that the 15% WD-UNet outperforms supervised models in DSC, TD, precision and IoU metrics, achieving scores of 0.926, 0.915, 0.884 and 0.865, respectively. In the BD metrics, where the difference is less pronounced, it attains scores of 0.857.

In Table II and Figure 6 an interesting trend becomes evident as the proportion of training data increases from 15% to 75% (n =152 to n=762). Initially, the model achieves a high DSC value of 0.926 at 15%, then this metric increases to 0.930 at 35%. Intriguingly, the DSC exhibits a decrease from 35% to 75%, decreases to 0.912. The trends in precision and IoU metrics align with those observed in DSC. Precision demonstrates a performance of 0.884 at 15% training data (n=152), which increases to 0.895 at 35% but then declines to 0.852 as the training data expands from 35% to 75% (n=356 to n=762). Similarly, IoU starts at 0.865 with 15% training data, rises to 0.871 at 35%, and subsequently decreases to 0.861 when the training data is increased from 35% to 75%. Conversely, the trends in BD and TD metrics are entirely opposite to those observed in DSC. BD starts at 0.857 with 15% training data, increases to 0.868 at 35%, and further ascends to 0.920 as the training data expands from 35% to 75%. In the case of TD, it initially registers at 0.915 with 15% training data, decreases to 0.897 at 35%, but then rises to 0.945 when the training data is increased from 35% to 75%. Since using only 15% of the training data can achieve good results, domain experts can annotate a small dataset of just 15% for training. However, to ensure the model does not overfit, the small dataset should theoretically ensure diversity in its contained samples. For instance, in lung airway segmentation, this could include training samples from various ages, genders, presence of lung disease, etc. Simultaneously, compared to the 35% WD-UNet, the 75% WD-UNet shows a slight decline in performance metrics such as DSC, IoU, and Precision. However, it exhibits a significant improvement with $P<0.001$ in both BD and TD. As observed in Figure 8, this enhancement is attributed to the model's ability to predict the green-colored distal branches of the airways, which are conventionally considered false positives. Yet, it can also be interpreted as the model identifying subtle distal airway branches that are not readily visible to domain experts. These findings lead to a reasonable inference: Human-Computer Interaction-based deep learning methods perform remarkably well with smaller datasets, experience a slight performance drop of DSC, Precision and IoU as diversity increases (potentially due to out-of-scope samples), and ultimately stabilize, achieving improved the evaluation metrics thanks to the robust generalization capability of the Human-Computer Interaction-based deep learning model. Additionally, the better performance of the 15% WD-UNet compared to the 35% WD-UNet in TD should not be misconstrued as overfitting of the model. From box plots (Figure 6), it is observed that 35% WD and 55% WD show no significant differences in the metrics of Intersection over Union (IoU), Dice, and Precision (P=0.058, P=0.079, P=0.063). Similarly, 15% WD and 55% WD demonstrate no significant discrepancies with P=0.0578 and P=0.0511 in the metrics of Tree Detection (TD) and Branch Detection (BD). This suggests that, in terms of similarity, the performance of 35% WD can nearly match that of 55% WD. In terms of the predictive performance on the number and length of airway branches, 15% WD is capable of identifying some of the finer airway branches at a level comparable to 75% WD; however, it is not sufficiently accurate and may include some errors (False positives).

TABLE II. *Evaluation metrics for comparing the performance of airway segmentation between supervised and semi-supervised models with varying training data ratios. Evaluation metrics include Dice Similarity Coefficient (DSC), branch score (BD), tree detected ratio (TD), and Intersection over Union (IoU). In the table, we highlight in bold the model performance metrics that merit specific attention and in-depth discussion within the paper. *UNet is enhanced with the addition of Branch Loss and Central Line Loss, which are more specifically tailored for airway segmentation.*

| Evaluation Metrics | Our Models | | | | | | |
|---|---|---|---|---|---|---|---|
| | Supervised model | | Semi-supervised models (with human-in-the-loop integrated) | | | | |
| | UNet (100% training data) | CEUNet (100% training data) | Proposed WD-UNet (15% training data) | Proposed WD-UNet (35% training data) | Proposed WD-UNet (55% training data) | Proposed WD-UNet (75% training data) | Proposed WD-UNet (100% training data) |
| DSC | 0.872±0.029 | 0.868±0.048 | 0.926±0.037 | **0.930±0.038** | 0.924±0.037 | 0.912±0.035 | 0.931±0.034 |
| Precision | 0.861±0.040 | 0.875±0.033 | 0.884±0.056 | **0.895±0.047** | 0.885±0.046 | 0.852±0.053 | 0.902±0.043 |
| TD | 0.896±0.121 | 0.859±0.171 | 0.915±0.085 | 0.897±0.104 | 0.896±0.110 | **0.945±0.078** | 0.873±0.071 |
| BD | 0.876±0.151 | 0.834±0.197 | 0.857±0.104 | 0.868±0.130 | 0.880±0.140 | **0.920±0.140** | 0.858±0.090 |
| IoU | 0.773±0.044 | 0.770±0.066 | 0.865±0.061 | **0.871±0.063** | 0.861±0.060 | 0.840±0.057 | 0.886±0.046 |

This interpretation is informed by the distinct nature of our training and test datasets: the training set consists of normal pulmonary airway CT images, while the test set is composed of CT images of pulmonary organs from patients with fibrosis and COVID-19, representing two completely non-overlapping datasets.

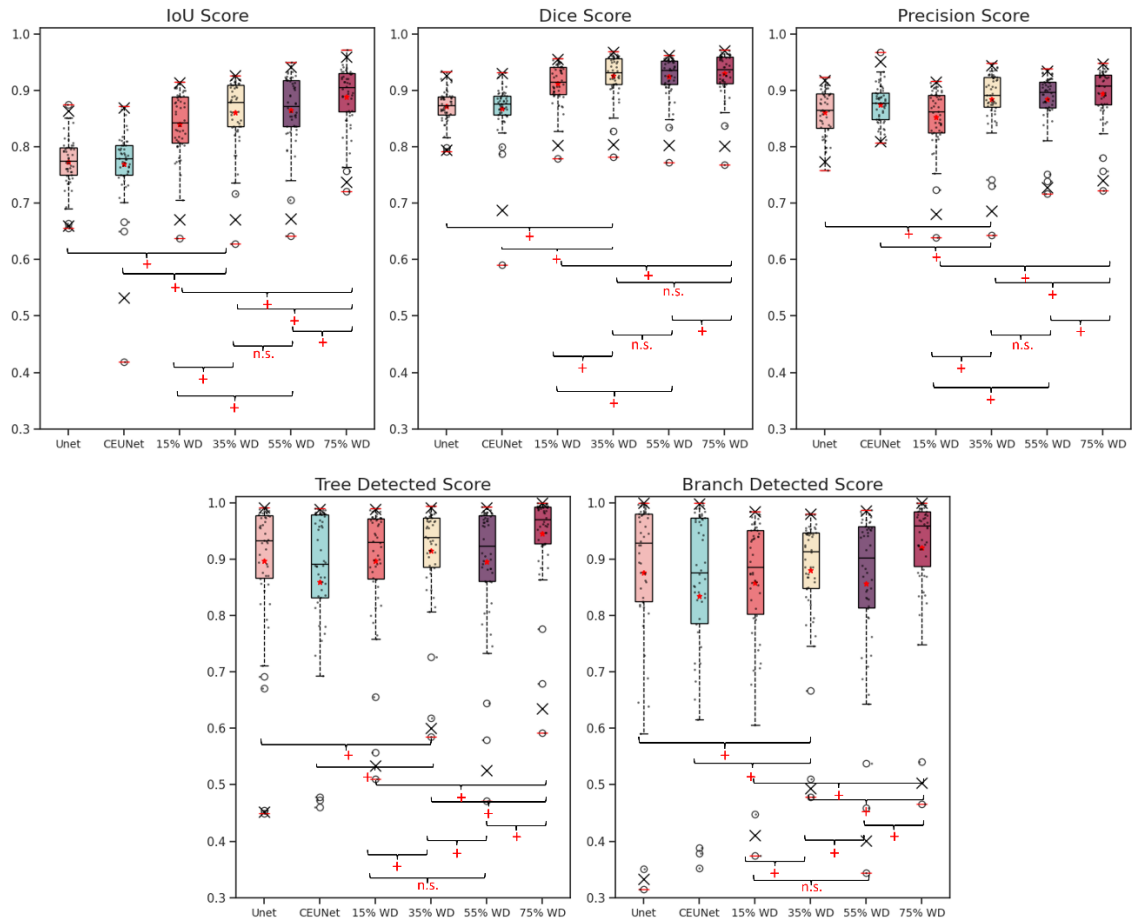

*Figure 6. Evaluation metrics for comparing the performance of airway segmentation between supervised and HCI models (WD-UNet) with varying training data ratios(15%, 35%, 55%, 75%, n= 152, 356, 559, 762 patches) and central line correction. * This Figure 6 is*

*associated with Table II. (Box range: interquartile Range; "-": minimum/maximum; "x": 1% and 99% confidence interval; "-": median; "\*": mean; "o": outliers; ".": scatter plot of the data distribution; "+": p<0.05; "n.s.": no significant difference between two groups.). Statistical significances were given by Paired Samples t-Test.*

Table III and Figure 7 presents the outcomes of our experiments in which LC-UNet, UUNet, RS-UNet, and our proposed WD-UNet model were trained using a 35% portion of the available training data (n=356) with central line correction. A comprehensive assessment was conducted to juxtapose these models against the two supervised models. The results of these experiments demonstrated the superior performance of the WD-UNet model when compared to the other three Human-Computer Interaction-based deep learning models, as evidenced by its excellence across all five performance metrics. It is evident that WD-UNet outperforms LC-UNet, UUNet, and RS-UNet across every evaluation metric, positioning it as the optimal model among the group. When comparing the performance of LC-UNet, UUNet, and RS-UNet to supervised models, they are slightly lower than UNet's TD and BD of 0.896 and 0.876. However, these models exceed the supervised models in DSC, achieving respective scores of 0.921, 0.921, and 0.913. In precision metrics, LC-UNet and UUNet surpass the supervised models with scores of 0.887 and 0.889. Similarly, in IoU, they also notably outperform the supervised models, achieving scores of 0.874, 0.870, and 0.842. From the box plots in Figure 7, it is evident that WD-UNet, except for IoU, where it shows no significant difference from UUnet (P=0.069), exhibits significant differences in the remaining four metrics compared to the other three HCI models and two supervised models (p<0.05). Simultaneously, LC-UNet and UUnet show no significant differences in IoU, Dice, and Precision metrics (P=0.113, P=0.05, P=0.124), and LC-UNet and RS-UNet exhibit no significant differences in Branch Detection (BD) and Tree Detection (TD) metrics (P=0.770, P=0.079).

TABLE III. Evaluation metrics of the airway segmentation results between 2 supervised models and 4 semi-supervised models. * "BF" represents the prediction results obtained before the inclusion of expert manual corrections to the centerline, while "Aft." denotes the prediction results obtained after incorporating the expert manual corrections to the centerline.

| Evaluation Metrics | Our Models | | | | | |
|---|---|---|---|---|---|---|
| | Supervised model | | Semi-supervised models (with human-in-the-loop integrated) | | | |
| | UNet (100% training data) | CEUNet (100% training data) | LC-UNet (35% training data) | UUNet (35% training data) | RS-UNet (35% training data) | Proposed WD-UNet (35% training data) |
| DSC (BF) | 0.872±0.029 | 0.868±0.048 | 0.921 ±0.033 | 0.921±0.036 | 0.913±0.035 | 0.930 ±0.038 |
| Precision (BF) | 0.861±0.040 | 0.875±0.033 | 0.887 ±0.048 | 0.889±0.041 | 0.861 ±0.046 | 0.895± 0.047 |
| TD (BF) | 0.896±0.121 | 0.859±0.171 | 0.884± 0.099 | 0.890±0.111 | 0.889± 0.098 | 0.897 ±0.104 |
| BD (BF) | 0.876±0.151 | 0.834±0.197 | 0.859± 0.130 | 0.852±0.136 | 0.855± 0.129 | 0.868± 0.130 |
| IoU (BF) | 0.773±0.044 | 0.770±0.066 | 0.863 ±0.055 | 0.863±0.059 | 0.839±0.058 | 0.871 ±0.063 |
| DSC (Aft.) | 0.893±0.028 | 0.879±0.024 | 0.924 ±0.028 | 0.928±0.032 | 0.919±0.032 | **0.939±0.031** |
| Precision (Aft.) | 0.869±0.045 | 0.878±0.042 | 0.912 ±0.035 | 0.908±0.037 | 0.900 ±0.034 | **0.930±0.042** |
| TD (Aft.) | 0.885±0.047 | 0.848±0.048 | 0.870± 0.093 | 0.874±0.098 | 0.860± 0.119 | **0.870±0.105** |
| BD (Aft.) | 0.864±0.078 | 0.822±0.098 | 0.803± 0.124 | 0.804±0.124 | 0.790± 0.013 | **0.831±0.113** |
| IoU (Aft.) | 0.817±0.055 | 0.805±0.056 | 0.874 ±0.046 | 0.870±0.054 | 0.842±0.065 | **0.891±0.059** |

When radiology experts annotated the selected unlabeled samples and made corrections to the tracheal centerline, the model's predictive performance underwent great changes. Comparing the model's performance before and after the addition of the centerline as a training loss, notable improvements were observed in metrics such as DSC, Precision, and IoU, while TD and BD values exhibited slight declines. Observations from TABLE III reveal that when the central lines modified by domain experts are incorporated as part of the training loss, there is a notable increase in the metrics of DSC, Precision, and IoU, ascending to 0.939, 0.930, and 0.891, respectively. However, there is a decline in TD and BD, which decrease to 0.870 and 0.831, respectively. However, it's important to note that the reductions in TD and BD values are not necessarily negative outcomes. The decrease in BD values can be attributed to the refinement and revision of the centerline by human experts, resulting in a reduction in the calculated branch count, effectively eliminating erroneous tracheal segments. The decrease in TD values is due to the removal of extraneous looping sections of the trachea by domain experts, leading to a reduction in the overall length of the predicted trachea. A paired samples Wilcoxon signed-rank test in conducted on the Branch Ratio results of the WD-UNet model, pre- and post-application of the central line, yielding a p-value of 1e-7 ($P<0.001$). This result substantiates that the incorporation of central lines annotated by domain experts impacts with statistical significance on the model's performance.

In several studies, severity of traction bronchiectasis, which represents abnormal dilation of the airways caused by the surrounding fibrotic lung tissue, has been reported as a strong predictor of mortality in several fibrotic lung disease subsets including idiopathic fibrotic lung disease, connective tissue disease (CTD) related fibrotic lung disease and chronic fibrotic hypersensitivity pneumonitis[53]. Therefore, abnormally dilated airways can serve as a biomarker for diagnosing fibrotic lung diseases. Radiologists, when quantifying these dilated airways, need to rely on precise airway branches. However, the central lines extracted using the skimage.morphology package are inaccurate, leading to imprecise branch levels and nodes in the airway branches. For this reason, these changes collectively contribute to a more accurate and clinically meaningful prediction of the airway structure.

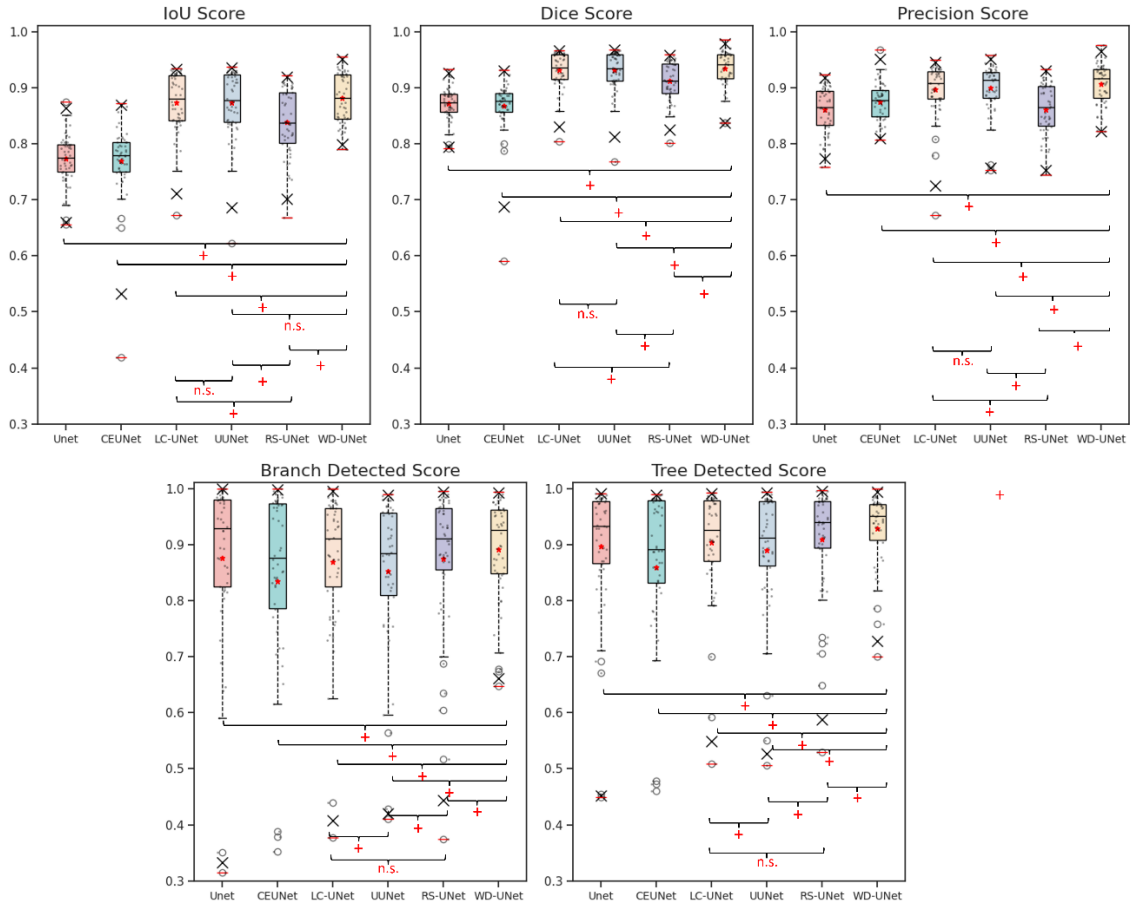

*Figure 7. Evaluation metrics of the airway segmentation results between 2 supervised models and 4 HCI models with central line correction. *This Figure 7 is associated with Table III. (Box range: interquartile Range; "-": minimum/maximum; "x": 1% and 99% confidence interval; "-": median; "*": mean; "o": outliers; ".": scatter plot of the data distribution; "+": p<0.05; "n.s.": no significant difference between two groups.). Statistical significances were given by Paired Samples t-Test.*

In Figure 8, radiology experts annotate and correct the centerlines after each round, which illustrates the enhancement effects on the models facilitated by the proposed Human-Computer Interaction (HCI) learning. The performance enhancements are demonstrated through five evaluation metrics: IoU, Precision, Dice, TD, and BD. Each round represents a cycle of expert intervention relative to the model's predictive performance before the intervention. Within Figure 8, each round comprises 10 epochs, and after every HCI iteration round, 10 samples are selected for expert annotation and centerline correction. Part (a) illustrates the proposed WD_UNet model trained with only 15% of the training data (n=152), showing a steady improvement in model performance with increasing rounds of expert involvement, peaking at the 12th round (IoU: 0.878, Precision: 0.916, Dice: 0.932, BD: 0.460, TD: 0.922), followed by fluctuations and eventual stabilization, achieving best observed performance in the final round (IoU: 0.883, Precision: 0.926, Dice: 0.935, BD: 0.460, TD: 0.920). Part (b) features the proposed WD_UNet model trained with only 35% of the training data (n=356), also exhibiting steady performance enhancements with expert participation, reaching an initial peak at the 5th round (IoU: 0.869, Precision: 0.916, Dice: 0.926, BD: 0.456, TD: 0.890), and fluctuating towards stability, with the ultimate best performance in the last round (IoU: 0.888, Precision: 0.926, Dice: 0.938, BD: 0.463, TD: 0.921). Through the line charts of evaluation metrics, it is observed that starting with 35% training data (n=365) converges more quickly and stabilizes faster than starting with 15% training data (n=152), achieving a slightly better model performance in the final round, albeit with minimal difference. Therefore, if annotation time is limited, choosing to train the model with only 15% of the annotated data could achieve near-comparable results to using 35% of the data. Figure 8 validates the models post each proposed HCI round training, based on 3D patches of size [128, 96, 144], where BD signifies the intersection of ground truth and predicted airways, focusing not only on airway boundaries but also on the overall structure, hence the slightly lower BD values. Meanwhile, TD, representing the intersection of predicted airways with the centerline, shows higher values. Combining this with the other four metrics suggests that the high TD values may be due to complete coverage of the centerline, although accurate prediction of the complete airway boundaries should be assessed with the other four metrics as well.

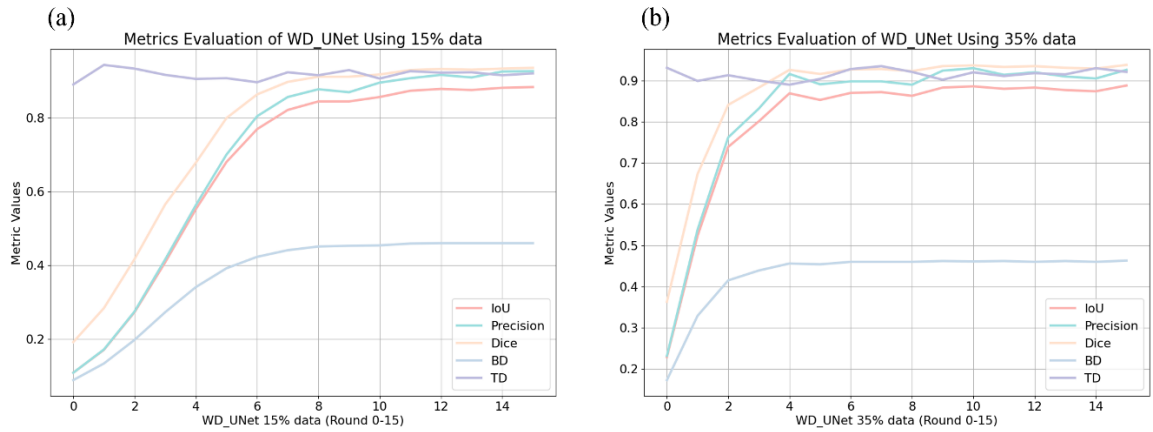

*Figure 8. Evaluation Metrics Improvement on the Proposed WD_UNet Model **After Each Round Annotated by Radiologists**. *There's totally 15 Rounds of HCI, each round contains once experts' manual annotation. Curved lines of evaluation metrics are used to validate the impact of each expert intervention on the HCI models, including the trends of improvement and when the model performance ceases to improve.*

**4.3 Visualization Findings**

Figure 9 delineates the performance of WD-UNet trained with data ranging from 15% to 75% (n=152 to n= 762), as well as the performance of all proposed semi-supervised models. As the training proportion increases from 15% to 75%, the model, while maintaining accuracy in DSC, shifts more focus towards predicting the distal ends of the airways. The increasing green false positive parts, evident with higher training proportions, can be conceptualized as the subtle distal airway branches that are not visually discernible to domain experts. Among the four proposed semi-supervised models, WD-UNet exhibits the best performance. This is attributed to its ability to minimize the blue areas (false negatives), while appropriately and logically predicting finer airway branches at suitable distal ends. It is also observable that WD-UNet, across training proportions from 15% to 75% (n=152 to n=762), consistently achieves higher DSC scores than RS-UNet, UUNet, and LC-UNet, with respective scores of 0.800, 0.831, 0.867, and 0.871. Additionally, our experts randomly selected 10 test samples for visualization and manual counting. It was noted that in a random instance of visualization results (Figure 9), WD-UNet, when using between 15% to 75% of the training data (n=152 to n=762), exhibits fewer blue airway areas (indicative of false negatives) compared to RS-UNet, UUNet, and LC-UNet. However, this may only represent the predictive results for a portion of the samples. Therefore, a comprehensive discussion should also incorporate the five evaluation metrics associated with the sample set. This finding further underscores the superior capability of WD-UNet in accurately capturing airway structures, especially in reducing the instances of missed detections.

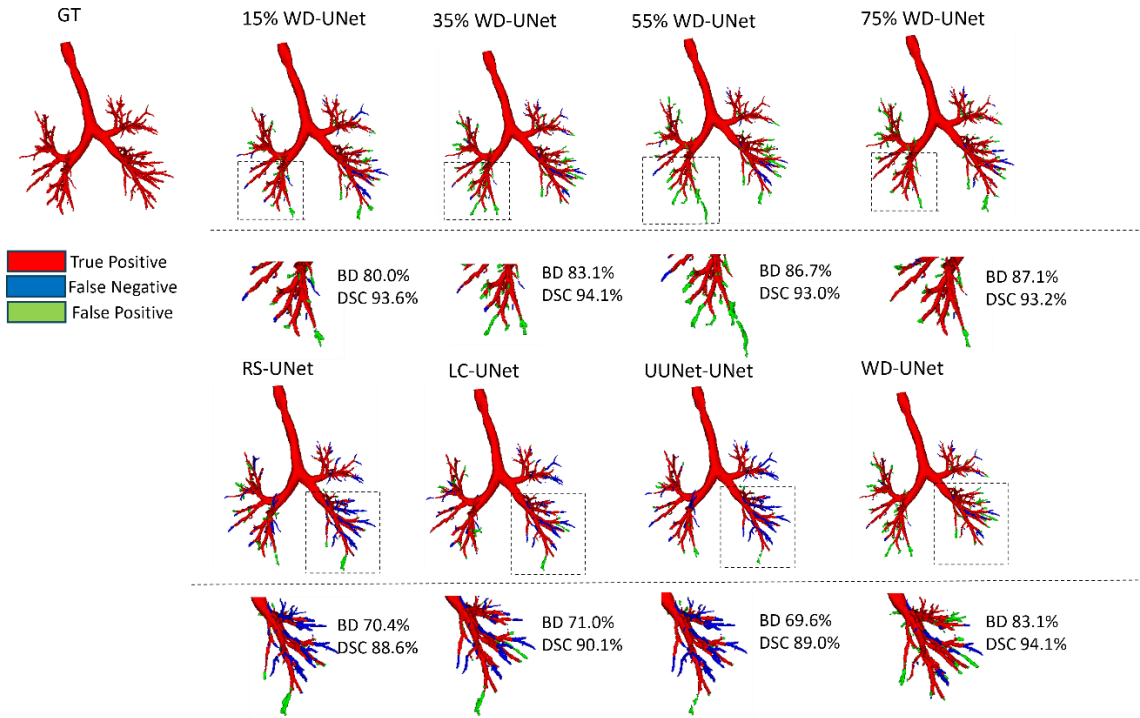

*Figure 9. The comparison between 15%-75% WD-UNet and all semi-supervised models.*

When we employing a mere 15% of the training data (n=152), the graphical representation in Figure 10 reveals that WD-UNet exhibits a relatively weaker predictive capacity for the distal trachea in comparison to CEUNet, with more noticeable false-negative regions, but it is higher in performance to UNet. Based on the visual inspection of Figure 10, it's observable that CEUNet and WD-UNet have similar false positive/false negative areas. Also, by comparing the values in Figure 10 for the four evaluation metrics – BD and DSC - which are closely matched at 0.873/0.861(CEUNet), 0.855/0.904(15% WD-UNet), respectively. However, as the proportion of training data increases to 35% (n=356), the predictive capability of the WD-UNet model greatly improves, approaching that of UNet with fewer false-negative regions. The green-shaded regions within the figures denote false-positive areas, suggesting that although some prediction errors may be present, they primarily highlight the model's proficiency in detecting subtle tracheal branches in the distal region, which may not be discernible even to expert radiologists.

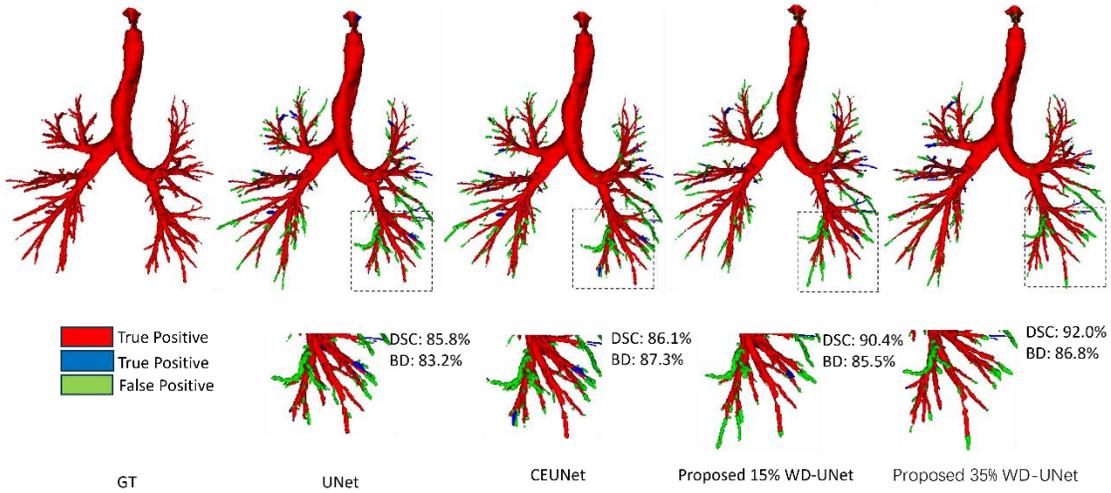

*Figure 10. Visualized prediction results of 2 pulmonary airways cases while use only 15% and 35% training data (n=152 and n=356). * The blue and green markings represent false negative and false positive, respectively.*

When comparing WD-UNet to supervised models, a visual inspection reveals that WD-UNet outperforms UNet and closely approaches CEUNet in terms of performance when utilizing only 15% of the training data (n=152). However, when comparing it to the other three Human-Computer Interaction-based deep learning models, it becomes evident that the prediction quality of distal fine branches is significantly better than the other three Human-Computer Interaction-based deep learning models, as shown in Figure 11. Furthermore, when considering the aspect of whether the central line has been optimized, it is found that training Human-Computer Interaction-based deep learning models with expert-optimized central lines leads to a notable enhancement in the prediction performance for all Human-Computer Interaction-based deep learning models. This improvement is most prominently observed in the reduction of False Negatives and False Positive, represented by the blue and green tracheal regions. The optimized prediction results for False Negatives and False Positives have been highlighted within black boxes in Figure 11.

The four proposed Human-Computer Interaction-based deep learning models have demonstrated remarkable performance, even when trained with only 15% of the available data (n=152). Particularly noteworthy is the consistent reduction in false negatives exhibited by WD-UNet as the proportion of training data increases, indicative of its enhanced ability to predict intricate branches in the distal trachea. When translated into clinical applications, WD-UNet presents a notable advantage due to its capacity to dynamically balance performance and cost by optimizing the trade-off between evaluation metrics and the quantity of training data required. Furthermore, WD-UNet showcases its superiority when compared to other contemporary deep learning methods replicated in this investigation (refer to TABLE III and Figure 7), underscoring its effectiveness in diversity prediction and model stability.

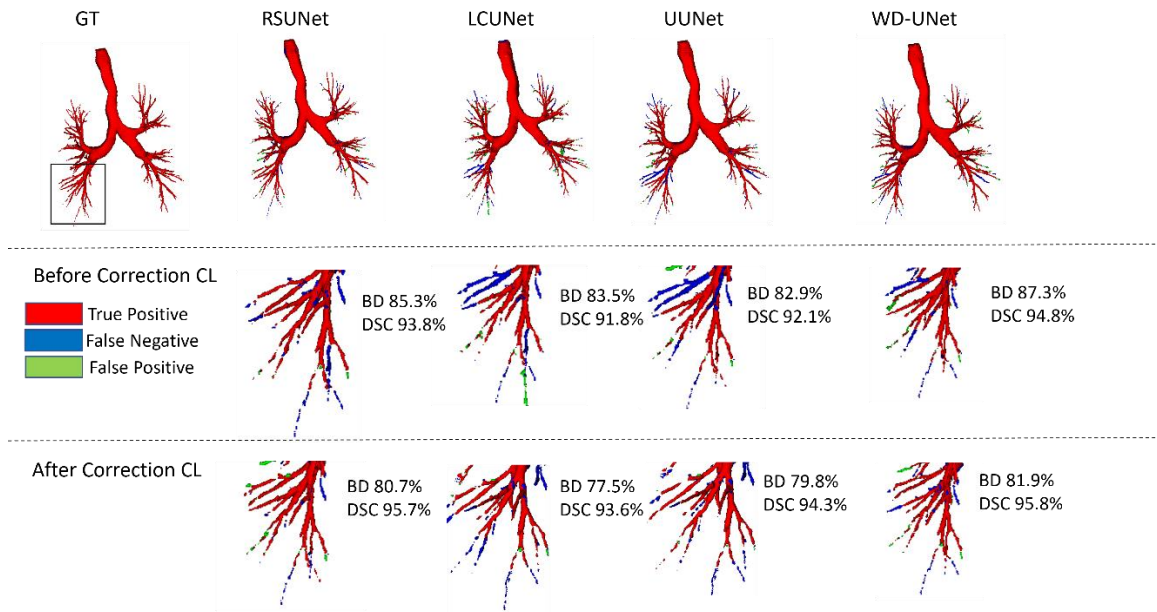

*Figure 11. A visual comparison of the prediction results from the four Human-Computer Interaction-based deep learning models. For the same test sample, it includes a comparison before and after adding the corrected central line. *CL represents the central line.*

## 5 Conclusion

In summary, our Human-Computer Interaction-based learning models comprises two distinct modules: lung airway segmentation and central line correction, which leverage the expertise of radiologists and their domain knowledge. The segmentation module is a collaborative effort involving radiological experts and is grounded in the Human-Computer Interaction-based deep learning algorithms. Remarkably, even with as little as 15% - 35% of the training data (n=152 to n=356), our proposed segmentation model attains evaluation metrics that are comparable to those of fully automatic supervised models, and it even surpasses them in certain metrics.

The development of a pseudo mask for segmenting airways, achieved by applying AI methods to centerline annotations from radiology experts, highlights the effectiveness of Human-Computer Interaction. This exemplifies that AI technologies are not solely designed in isolation but can be cultivated and refined through the collaborative synergy of Human-Computer Interaction-based learning models.

However, the proposed WD-UNet also has limitations, due to its more complex model structure. It incorporates not only a segmentation network but also a feature extractor and a discriminator, leading to longer training times compared to other semi-supervised and supervised models. In the future, we will focus on reducing model parameters and simplifying the algorithmic structure to optimize its performance.

**Acknowledgement**

This study was supported in part by the ERC IMI (101005122), the H2020 (952172), the MRC (MC/PC/21013), the Royal Society (IEC\NSFC\211235), the NVIDIA Academic Hardware Grant Program, the SABER project supported by Boehringer Ingelheim Ltd, NIHR Imperial Biomedical Research Centre (RDA01), Wellcome Leap Dynamic Resilience, UKRI guarantee funding for Horizon Europe MSCA Postdoctoral Fellowships (EP/Z002206/1), and the UKRI Future Leaders Fellowship (MR/V023799/1). Send correspondence to G. Yang g.yang@imperial.ac.uk.